\documentclass[12pt]{iopart}
\usepackage{graphicx}

\begin{document}

\title[Quantum rings with defects in a magnetic field]{Electron states in quantum rings with defects under axial or in-plane magnetic fields}

\author{J Planelles$^1$, F Rajadell$^1$ and JI Climente$^{2,1}$} 
\address{$^1$ Departament de Qu\'{\i}mica F\'{\i}sica, Universitat Jaume I, Box 224, E-12080 Castell\'o, Spain}
\address{$^2$ S3 CNR-INFM, Via Campi 213/A, 41100 Modena, Italy}

\ead{josep.planelles@qfa.uji.es}

\date{\today}

\begin{abstract}
A comprehensive study of anisotropic quantum rings, QRs, subject to axial and in-plane magnetic field, both aligned and transverse to the anisotropy direction, is carried out. Elliptical QRs for a wide range of eccentricity values and also perfectly circular QRs including one or more barriers disturbing persistent QR current are considered. These models mimic anisotropic geometry deformations and mass diffusion occuring in the QR fabrication process. Symmetry considerations and simplified analytical models supply physical insight into the obtained numerical results. Our study demonstrates that, except for unusual extremely large eccentricities, QR geometry deformations only appreciably influence a few low-lying states, while the effect of barriers disturbing the QR persistent current is stronger and affects all studied states to a similar extent.
We also show that the response of the electron states to in-plane magnetic fields provides accurate
information on the structural anisotropy.
\end{abstract}

\submitto{\NT}
\maketitle
\section{Introduction}
Ring shaped semiconductor nanostructures, called nanorings or quantum rings (QRs), have gathered large attention in the last two decades,\cite{ButtikerPL,AronovRMP,ChakraPRB,WendlerPSSb,FuhrerNAT,PlanellesPRB,DasilvaPRB,MiquelPRB} mainly due to the magnetic properties they can display, related to their double connected topology. In particular, small QRs are the best candidates to show pure quantum effects as e.g. the Aharonov-Bohm (AB) effect,\cite{ABhom} because they are in the scattering-free and few-particle limit.\cite{LorkePRL}\\

In addition to the theoretical or fundamental interest of these nanostructures, QRs, like others-shape nanostructures, have a highly relevant role in technological applications, as they may constitute the core of laser emitters,\cite{Bimberg,ParkJQE} storage devices,\cite{Yusa,Finley,Lundstrom} fluorescence markers,\cite{Bruchez} etc. \\

Most of theoretical studies, including sophisticated 3D models employed to describe experimental QRs magnetospectra,\cite{ClimentePRB,CCCC,CPRJPCM} assume perfect circular shape for QRs. However, although almost perfect circular or slightly oval shape QRs have been fabricated,\cite{GarciaAPL,LorkePRE} anisotropic samples of notably elongated QRs along a given direction have also been synthesized.\cite{Sormunen,RazAPL} We may consider perfect circular shape QRs as an idealization of grown solid-state rings, where more or less severe imperfections occur.\\

Theoretical modelizations devoted to analyze how defects influence the QR energy structure, in absence or in the presence of an axial magnetic field, have been carried out, including studies on eccentricity of the inner hole,\cite{BrunoPRB} ellipticity\cite{BermanPRB,MagarillJETP} and more general ring-deformations,\cite{GridinPRB} the presence of transverse barriers\cite{ZhuPRB} and doping with shallow donor impurities.\cite{DasilvaPRB,PanJPCM} Much less attention has been paid to the influence of in-plane or tilted magnetic field on the energy spectra of these nanosystems.\cite{planellesPE,ProcICNT} \\

In the present paper a comprehensive study of anisotropic QRs subject to axial and in-plane magnetic field, including both directions, along and transverse to the deformation, is carried out. In a first place, trying to mimic experimental geometry deformations,\cite{Sormunen,RazAPL} we deal with geometric anisotropy. Thus, we investigate the spectra of elliptical QRs, subject to homogeneous magnetic fields in the three abovementioned directions, vs. eccentricity. In a second instance, anisotropic mass diffusion is modeled by a perfect circular shape QR with a parallel double barrier, dividing it into two identical parts. Note that the parallel double-barrier structure was used in an experiment of the magneto-electric AB effect in metal rings.\cite{Oudenaarden} This system is also subject to an homogenous magnetic field in all three directions. A detailed study on oscillation modes in double-barrier QRs subject to an axial magnetic field has been recently reported\cite{ZhuPRB}. In order to go deeper on the understanding of this issue, a set of calculations of QRs pierced by an axial magnetic field and including no transverse barriers ($C_{\infty}$ symmetry), four barriers ($C_4$), three barriers ($C_3$), two parallel barriers ($C_2$), two non-parallel barriers and a single barrier (no symmetry), is carried out. The aim of this section is to relate the obtained results to symmetry considerations, which allows for a very simple analysis of oscillation modes. Finally, random imperfections are simulated by including a single transverse and incomplete barrier. We study the evolution of energy magnetospectra vs. the length of this small barrier up to the case it completely cuts the QR. The paper is organized as follows, next section is devoted to theoretical considerations on the Hamiltonian and magnetic field gauges employed. It is followed by a section on results and discussions, which is in turn organized in several subsections devoted to the different case-studies. Finally a concluding remarks section ends the paper.

\section{Theory}

In this paper we study single-electron energy levels of semiconductor anisotropic QRs 
subject to both, axial and in-plane magnetic fields. Since the vertical confinement in these structures is 
much stronger than the lateral one,\cite{LorkePRE,Sormunen,RazAPL}  we employ a two-dimensional Hamiltonian 
to describe the low-lying energy levels. Similar models have proved successful in explaining the fundamental physics 
of QRs in the presence of axial magnetic fields.\cite{LorkePRL,Chakra_book,BayerPRL,bidimen}\\

Within the effective mass and envelope function approximations, this Hamiltonian for a system in the $(x,y)$ plane reads, in atomic units, 

\begin{equation}
\label{eq1}
H=\frac{1}{2 m^*} (\mathbf{p}+\mathbf{A})^2+V(x,y),
\end{equation}

\noindent where $m^*$ stands for the electron effective mass, $\mathbf{A}$ is the vector potential and $V(x,y)$ represents 
a finite scalar potential which confines the electron within an annular finite region of the space (QR). This confining potential 
is then defined as:

\begin{equation}
\label{eqV}
V(x,y)=\left\{ \begin{array}{ll}
0 & {\rm if} \; (x,y) \in {\rm QR}\\
V_c & {\rm if}  (x,y) \in {\rm surrounding \;\; matrix,} \end{array} \right.
\end{equation}

\noindent where $V_c$ stands for the heterostructure band-offset.\\

We employ the so-called Coulomb gauge, defined by the condition $\nabla \cdot \mathbf{A}=0$. Then, an axial magnetic field  
$\mathbf{B_{axial}}=(0,0,B)$ may be derived from the vector potential\cite{note1} $\mathbf{A_{axial}}=\frac{1}{2}(-y,x,0) B$.
The in-plane magnetic field, applied along the $x$-axis, $\mathbf{B_{in}}=(B,0,0)$ may be in turn derived from the vector potential\cite{note2}$\mathbf{A_{in}}=(0,0,y) B$.\\

Therefore, the Hamiltonian in the presence of an axial magnetic field finally reads:

\begin{eqnarray}
\label{eq2}
H&=&\frac{1}{2 m^*} (\hat p_x^2+\hat p_y^2)+ \frac{B^2}{8 \; m^*} (x^2+y^2) - \nonumber \\ 
&& - i \frac{B}{2 \; m^*} (x \frac{\partial}{\partial y}-y \frac{\partial}{\partial x}) + V(x,y),    
\end{eqnarray}

\noindent while in the presence of in-plane magnetic applied along the $x$-axis, the Hamiltonian is of the form:

\begin{equation}
\label{eq3}
H=\frac{1}{2 m^*} (\hat p_x^2+\hat p_y^2)+ \frac{B^2 }{2 \; m^*} y^2 + V(x,y).
\end{equation}

It is worth pointing out that the two-dimensional approximation we use here is valid as long as the in-plane 
magnetic field term $2 B y \hat p_z$, which couples vertical and in-plane motions, is much smaller 
than the energy separation in the $z$-direction (see (4) in appendix of \cite{planellesPE}).
This condition holds for usual nanorings and moderate magnetic fields, owing to the strong vertical
confinement. As a consequence, an effective separation of variables is possible and only the
ground state in the $z$-direction is relevant for the energetically lower part of the electronic 
spectrum.  Still, in the case of thicker systems (or very weak confinement in the $y$ direction), the 
magnetic term $2 B y \hat p_z$ may significantly influence the low-lying part of the spectrum 
and the two-dimensional description would start failing.\\

The eigenvalue equations of the abovementioned Hamiltonians, (\ref{eq2}) and (\ref{eq3}), have been 
solved numerically using a finite-difference method on a two-dimensional grid ($x,y$) extended far beyond the QR limits. 
This discretization yields an eigenvalue problem of a huge asymmetric complex sparse matrix 
that has been solved in turn by employing the iterative Arnoldi factorization.\cite{arnoldi}

\section{Results and Discussion}
The QR system under consideration is the same we explored in previous papers where we focused on the effect of a tilted magnetic field on perfect circular QR\cite{planellesPE} and where we investigated the evolution of the energy spectra of a lateral double QR molecule when it dissociates subject to the influence of a magnetic field.\cite{ProcICNT} The present paper, as stated above, is devoted to study the role of imperfections. The perfect circular QR taken as reference is then a GaAs QR embedded in an Al$_{0.3}$Ga$_{0.7}$As matrix, the AlGaAs material acting as a barrier for the conduction band electrons confined in the GaAs ring. We use the same parameters employed in  \cite{planellesPE,ProcICNT}. Namely, an effective mass $m^*=$0.067, a band offset  $V_c=0.262$ eV, and, for the perfect circular QR reference, an inner radius $r_{in}=$12 nm and an outer one $r_{out}=$16 nm. A range of 0-20 T for the external magnetic field is considered. We have neglected the Zeeman term in (\ref{eq1}) as we focus on orbital effects of the magnetic field only. Indeed, the Zeeman splitting in the structures we study (of about 0.5 meV at 20T) is generally small as compared to the diamagnetic shifts.

\subsection{Elliptical QRs}
In this subsection we deal with geometry deformations.\cite{Sormunen,RazAPL} The considered QR is now elliptical. 
Ellipses with the same area (and then pierced by same magnetic flux at every strenght of the applied magnetic field) but with different eccentricities are considered. The shared area of these QRs is given by $A=\pi (r_{out}^2-r_{in}^2)$.
The inner/outer shorter axis $r_{in}^{(s)}/r_{out}^{(s)}$ of the ellipse is calculated from $r_{in}/r_{out}$ as $r_{in}^{(s)}=r_{in} \sqrt[4]{1-e^2}$ / $r_{out}^{(s)}=r_{out} \sqrt[4]{1-e^2}$. The inner/outer larger axis $r_{in}^{(l)}/r_{out}^{(l)}$ being related to the shorter axis $r_{in}^{(s)}/r_{out}^{(s)}$ by $r_{in}^{(s)}=r_{in}^{(l)} \sqrt{1-e^2}$ / $r_{out}^{(s)}=r_{out}^{(l)} \sqrt{1-e^2}$. The (2D) system is located in the $xy$ plane. Then, the axial magnetic field is aligned in the $z$ direction, and, as stated above, the in-plane magnetic field is applied along the $x$ axis. The larger axis of the ellipse, $R$, lies either along the $x$ (eccentricity and magnetic field aligned) or the $y$ axis (eccentricity transverse to the magnetic field).\\

\begin{figure}[p]
\begin{center}
\includegraphics[width=0.7\columnwidth,angle=270]{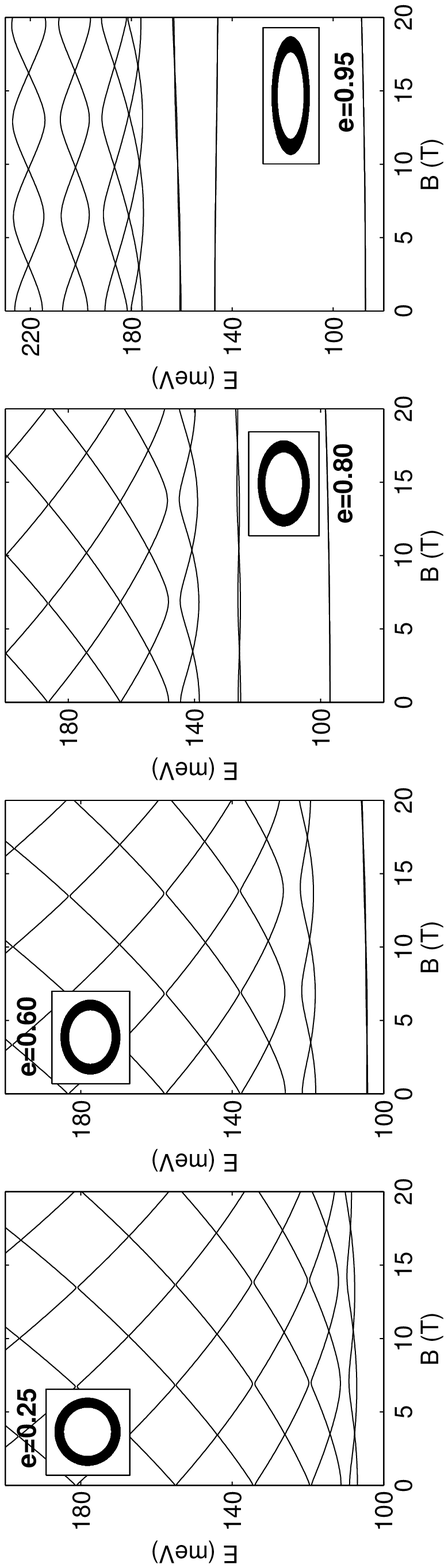}
\caption{Electron low-lying energy levels in QRs with eccentricity $e$, subject to axial magnetic fields.
The insets schematically depict the QRs shape.}\label{Fig1}
\end{center}
\end{figure}

Figure \ref{Fig1} illustrates the electron energy levels of a QR subject to an axial magnetic field for several eccentricities. In the presence of an axial magnetic field, going from circle to ellipse means reducing symmetry from $C_{\infty}$ to $C_2$ group. The last group has only two irreducible representations. Then, from symmetry considerations, one should expect an energy spectrum splitted into non-crossing pairs of states which in turn cross repeatedly as B increases (each pair of repeatedly crossing states containing one instance of each of the two $C_2$ symmetries). However, as it can be seen in figure \ref{Fig1}, 
this is only the case for severe ($e=0.95$) eccentricities.  For eccentricities smaller than 0.25 only a pair of low-lying states show anti-crossings and separate from the rest of energy levels. 
The remaining (excited) states almost do not feel the eccentricity. When eccentricity reaches 0.60, a couple of pairs result well separated from the rest.\cite{degen} When it reaches 0.8, three pairs separate and, as stated above, for eccentricities larger than 0.95, all of calculated states separate in non-crossing pairs of states which, in turn, crosses repeatedly as B increases.\cite{accidental} In other words, except for the lower part of the spectrum, non-severe eccentric elliptic QRs cannot be distinguished from perfect circular QRs, as far as the energy spectrum is concerned.\\

A last physical fact to be pointed out is the behavior of the two lowest-lying states. These states, in contrast to the other (excited) states, are very sensitive to eccentricity and separate from the rest of the spectrum, even for quite low values of the eccentricity. If eccentricity exceeds $e=0.6$ they become quasi-degenerate, each of them localizing the electronic density in either, the left or the right hand side region of the ellipse, around the larger axis $R$.\\

\begin{figure}[p]
\begin{center}
\includegraphics[width=0.7\columnwidth,angle=270]{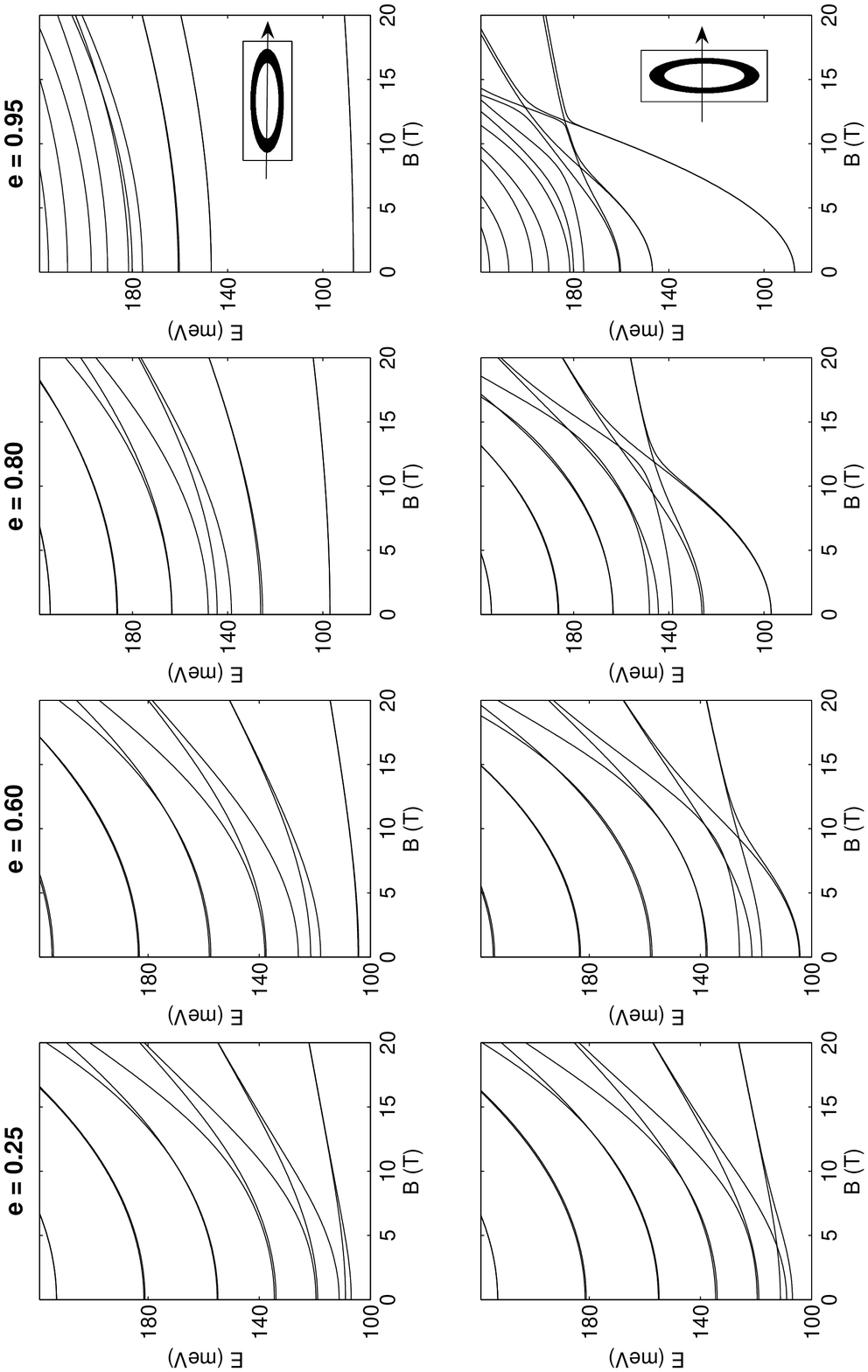}
\caption{Electron low-lying energy levels in QRs with eccentricity $e$, subject to in-plane magnetic fields.
The direction of the magnetic field with respect to the ring axes is represented by the arrow in the inset.
Upper (lower) panels: field parallel (perpendicular) to the axis of the ring larger radius.}\label{Fig2}
\end{center}
\end{figure}

Figure \ref{Fig2} shows the energy spectra vs. eccentricity of elliptical QRs subject to in-plane magnetic fields. Upper panels correspond to the larger axis $R$ of the ellipse oriented parallel to the magnetic field direction ($x$), while lower panels correspond to the axis $R$ perpendicular to the field direction. As it is revealed by (\ref{eq3}), the in-plane magnetic field just produces a parabolic confinement $B^2y^2/2m^*$ which tends to squeeze the electron wave function in the $y$ direction. The energy magneto-spectra are then completely different from the ones shown in figure \ref{Fig1} (axial field), the essential differences arising from the magnetic flux trapping in the QR.\cite{ABhom,OlariuRMP,PlanellesBGW} Figure \ref{Fig2} shows that, for a weak eccentricity, $e=0.25$, levels are arranged in pairs if the magnetic field is present. This effect, which was already observed in circular QRs\cite{planellesPE}, is due to the fact that the inner hole of the QR, together with a strong magnetic field which induces wave function compression in the $y$ direction, lead to the formation of quasi-degenerate double quantum well solutions along the $x$ axis. The magnetic field strenght required to reach the double well-like solutions is larger as the states are more excited, since their higher kinetic energy works more efficiently preventing the density localization in the double well.
This behavior holds in the presence of  more severe eccentricities, as can be seen in figure \ref{Fig2}, where, for large strenghts of the magnetic field, the arrangement in pairs of the levels becoming the even/odd solutions of a nearly one-dimensional double well, can be seen for eccentricities up to $e=0.8$. It should be stressed that what appears as a single low-lying line in the energy magneto-spectrum of aligned magnetic field and $e=0.60$ (figure \ref{Fig2}, second upper panel) is actually a superposition of two almost degenerate states. In a similar way, the two/three low-lying lines for $e=0.80/0.95$ are two/three pairs of quasi-degenerate states. A particular comment deserves the case  of aligned field and $e=0.95$. In this case no arrangement in pairs is produced by the magnetic field within the studied range of magnetic field. This fact parallels what can already be seen for quite excited states of $e=0.80$, where e.g. states 9-th and 10-th are still far from starting the energy convergency.

The differences among the energy spectra for the various studied cases mainly originate from the more or less severe splitting of the perfectly circular QR degenerate states produced by eccentricity.\\

\begin{figure}[p]
\begin{center}
\includegraphics[width=0.5\columnwidth,angle=270]{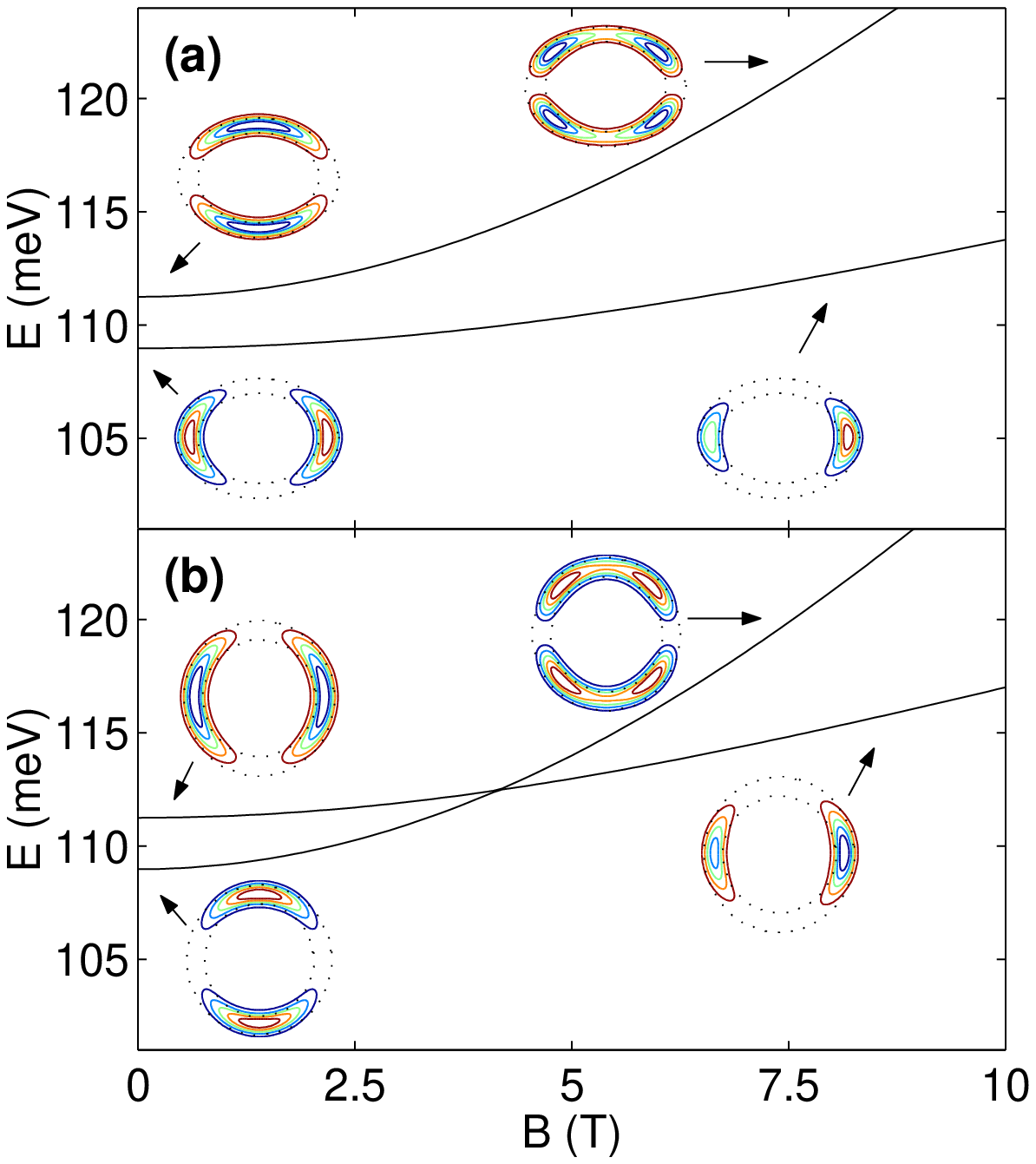}
\caption{(Color online). Energies and wave functions of the first and second excited states vs.~in-plane 
magnetic fields in a QR with $e=0.25$.  The orientation of the field in the upper (lower) panels is 
parallel (perpendicular) to the axis of the ring larger radius (see insets of figure \ref{Fig2}).
The contours represent the electron wave function at $B=0$ (left insets) and $B=10$ T (right insets),
with dotted lines showing the confinement potential profile.}\label{Fig3}
\end{center}
\end{figure}

If we now compare the effect of the in-plane magnetic field applied parallel or transverse to the ring larger axis $R$ (upper and lower panels of figure \ref{Fig2}) on the electron energy levels we see that, aside from qualitative features shared in both cases, such as the arrangement in pairs of levels at strong magnetic fields and non-severe eccentricities,  sizeable differences are also present. Namely (i) the strikingly different magnitude of the diamagnetic shifts of the low-lying levels with increasing eccentrity and (ii) the appearance of level crossings in the case of a field applied transverse to the ring larger axis.  The first difference is due to the fact that the lowest-lying eigenstates of elliptical QRs tend to localize in the sides of the rings (i.e. they are compressed around the $R$ axis).\cite{BermanPRB,GridinPRB} 
Hence, they barely feel a magnetic field parallel to $R$. On the contrary, when the field is perpendicular to $R$, it tends to localize the wave functions in the thinner arms of the elliptical QR, and this greatly unstabilizes the lowest-lying states.  
The second difference originates from the different location, relative to the direction of the field, of the wave-function nodes produced by eccentricity when splitting the degenerate ($\pm m$) states of a perfectly circular QR (here $m$ refers to the azimuthal angular momentum).  We illustrate this by investigating the second and third excited states of the QR with $e=0.25$ in figure \ref{Fig3}. In the absence of external fields, the first excited state (which, by analogy with that of the circular QR we may label $m=-1$) has the angular node transverse to the axis $R$, while the second excited one ($m=+1$) has the node parallel to $R$.  An external magnetic field applied along $R$ (upper panel) unstabilizes the $m=+1$ state while leaving the $m=-1$ almost unaltered. Thus, no crossing of levels occurs. However, the situation is reversed when the field is applied perpendicular to $R$ (lower panel), and this leads to the crossing between $m=\pm 1$ levels. The reason why in figure \ref{Fig2} only a single crossing is observed for $e=0.25$, while two of them are observed for $e=0.6$ and $e=0.8$, is that in the $e=0.25$ case the eccentrictiy only splits significantly states labeled as $m=\pm 1$ in perfect circular QR, while for $e=0.60$ and $e=0.80$ it splits both, $m=\pm 1$ and $m=\pm 2$ (see figure \ref{Fig1}). 
In the large eccentricity regime ($e=0.95$), even though several $\pm m$ states are non-degenerate at zero field (see the $e=0.95$ panel in figure \ref{Fig1}), only two crossings can be seen in the corresponding panel of figure \ref{Fig2}. This is because for states with many angular nodes, the unstabilization due to magnetic confinement is similar in each instance of the pair of states and cannot overcome, by far, the large splitting produced by eccentricity.
Finally, it is worth pointing out that the larger the eccentricity $e$, the larger the splitting produced by it at $B=0$T. 
Then, the crossings in the presence of  large eccentricities occur at magnetic fields larger than those for small 
eccentricities.

It follows from the above discussion that in-plane magnetic fields may provide valuable information on the
geometric deformation of QRs. Thus, if one measures e.g. the far-infrared magneto-spectrum of electrons in 
QRs subject to in-plane magnetic fields, changes in the excitation energies with the direction of the in-plane
field should could be seen as a signature of geometric anisotropy, the direction of ellipticity being that
of weakest dependence on the value of the field. Furthermore, the value of the magnetic field at which
the level crossings are observed for fields oriented transverse to the ring large axis give
quantitative estimates of the degree of ellipticity of the samples under study.\\

\subsection{Parallel double barrier QRs}

Here we mimic anisotropic diffusion. To this end, the reference perfectly circular QR is modified by enclosing a parallel double barrier lying along the QR diameter. This barrier, which has been selected to have a Gaussian profile, is defined by

\begin{equation}
\label{eq5}
V_b(\alpha,\beta)=\left\{ \begin{array}{ll}
V_c \, e^{-a \beta^2}& {\rm if} \;\;\;  -r_{out}< \alpha < -r_{in}  \\
                     & {\rm or}  \;\;\;\;\;\;\; r_{in}< \alpha < r_{out}\\
\\
0 & {\rm otherwise}, \end{array} \right.
\end{equation}
\noindent where $\alpha=x$ and $\beta=y$ for barriers aligned with the magnetic field, and opposite, i.e., $\alpha=y$ and $\beta=x$, if they are transverse.\\

The system is further perturbed by a constant magnetic field. As in the previous section, three directions for the magnetic field are explored. Namely, axial, in-plane aligned with and transverse to the double barrier.\\

\begin{figure}[p]
\begin{center}
\includegraphics[width=0.7\columnwidth,angle=270]{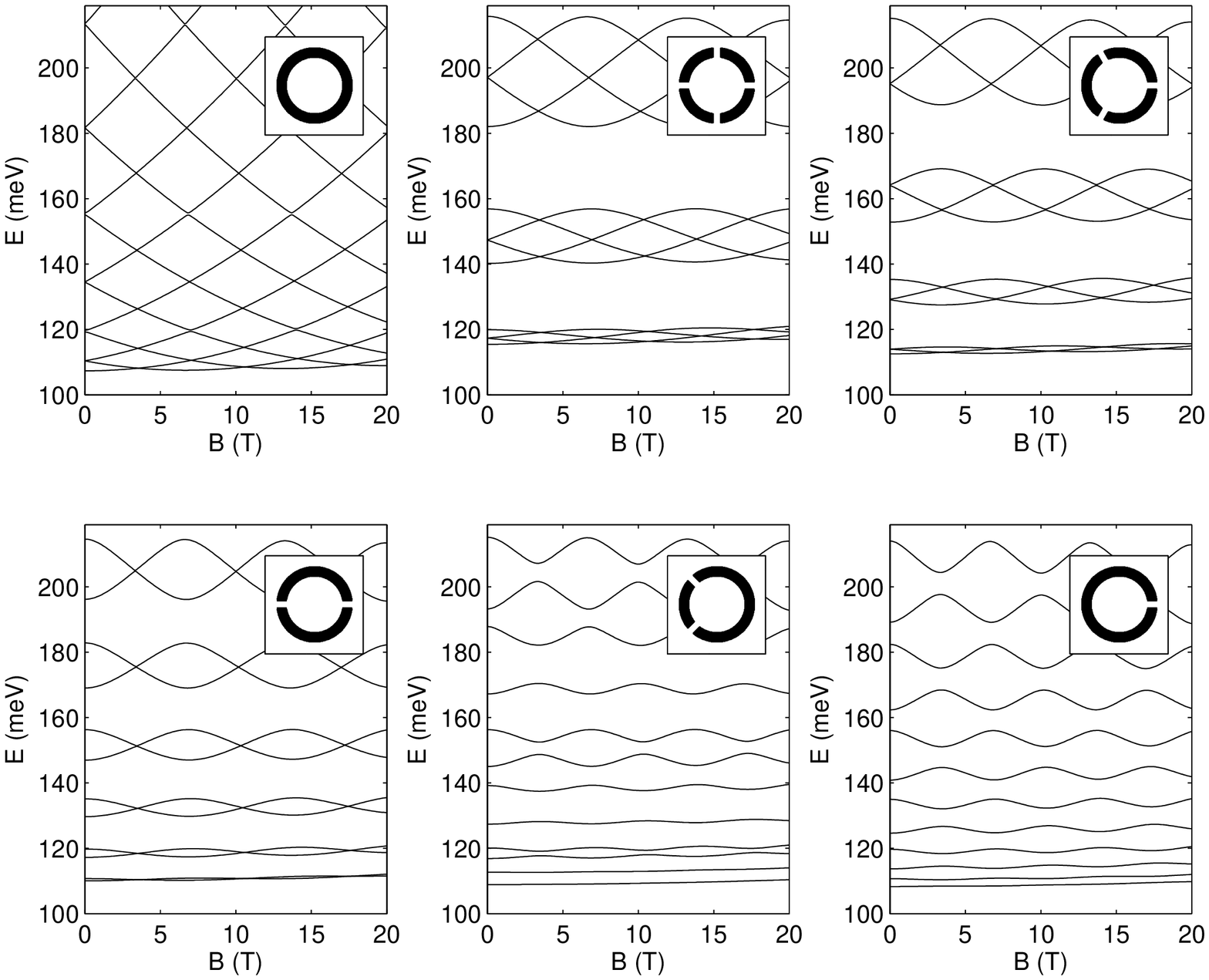}
\caption{Electron low-lying energy levels in QRs with potential barriers.
The insets depict the number and orientation of the barriers.}\label{Fig4}
\end{center}
\end{figure}

As stated above, a detailed study on the Aharonov-Bohm oscillation modes (i.e., modes modulated by an axial magnetic field) of parallel and non-parallel double-barrier QRs has been recently carried out.\cite{ZhuPRB} In that paper, a rather involved mathematical analysis allowed to relate oscillation modes of a barrier-less QR and QRs with a double barrier forming an angle. Particular mention is made, in the case of a parallel double barrier QR, of what the authors call ``occasional"  degeneracy at half flux unit. This study is doubtless interesting. However, group theory allows for a deeper, but all the same simpler, approach. It is carried out here. To make it easy, we present in figure \ref{Fig4} the low-lying electron energy levels of the abovementioned perfectly circular barrier-less QR pierced by an axial magnetic field, along with the corresponding energy levels of the QR modified by including symmetrically distributed (i) four Gaussian barriers (ii) three of them (iii) two of them (iv) two non-parallel barriers and (v) a single barrier (see insets). The relevant symmetry groups in the presence of an axial magnetic field are $C_{\infty}$, $C_4$, $C_3$, $C_2$, and $C_1$ (twice), respectively.\\

Figure \ref{Fig4} reveals that the spectra show crossings similar to those of the reference barrier-less QR. However, superimposed, there are anti-crossings coming from the particular symmetry of the system. Thus, we can see that the spectrum of the QR including four barriers ($C_4$) is splitted into non-crossing sets of four states. Within each set, the states cross repeatedly one another as $B$ increases. Every set contains one instance of each of the four $C_4$ symmetries, namely $A$, $B$, $E_+$, and $E_-$. In a similar way, the QR including three barriers ($C_3$) is splitted into non-crossing sets of three states, while the QR including a parallel double barrier ($C_2$) show pairs of states (one instance of each of the two $C_2$ symmetries) repeatedly crossing one another vs. $B$. These crossings are not really ``occasional" degeneracies: they take place in the case $C_n$ ($n=2,3,4,\dots$) simply because they take place also in the non-perturbed perfectly circular barrier-less QR. In this sense, they are as ``occasional" as all the Aharonov-Bohm crossings between different symmetry states of a perfectly circular barrier-less QR. Just on the contrary, the symmetry reduction $C_{\infty} \to C_n$ destroys many crossings, which turn into anti-crossings between states of the same symmetry. Indeed, in the cases of non-parallel double barrier and single barrier, symmetries have been completely removed ($C_1$) and all crossing turn into anti-crossings.\cite{e-e}

\begin{figure}[p]
\begin{center}
\includegraphics[width=0.7\columnwidth,angle=270]{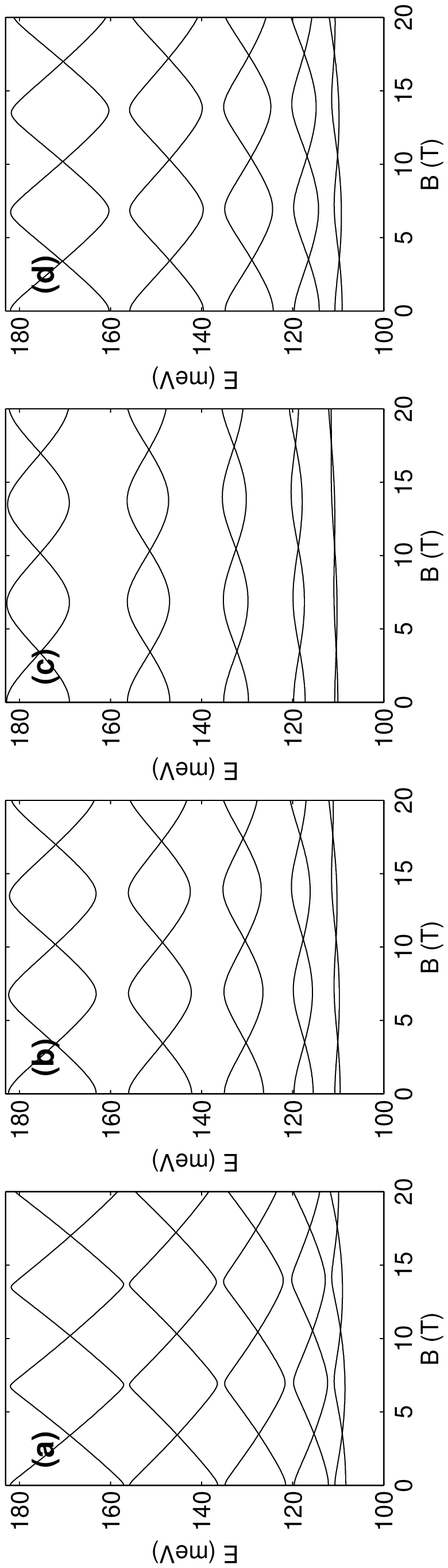}
\caption{Electron low-lying energy levels in a QR with parallel double barrier structures, 
subject to axial magnetic fields. The parameters of the barriers differ in each panel:
(a) width $a=0.003$, height $0.1\,V_c$;
(b) width $a=0.003$, height $0.5\,V_c$;
(c) width $a=0.003$, height $1.0\,V_c$;
(d) width $a=0.03$, height $0.1\,V_c$.
See (\ref{eq5}) for details about the barriers confinement potential.}\label{Fig5}
\end{center}
\end{figure}

Leaving this mathematical dissertation, let us come back to the simulation of an anisotropic diffusion by means of enclosing a parallel double barrier, (\ref{eq5}), in a barrier-less QR. At a first glance, one may think that eccentricity and parallel double barrier would yield the same qualitative physics. However, some interesting differences arise which deserve to be discussed. Thus, figure \ref{Fig5} displays the electron low-lying energy levels of a parallel double barrier QR pierced by an axial magnetic field. Several cases are accounted for. Panel (d) in figure \ref{Fig5} corresponds to a very narrow barrier ($a=0.03$, see (\ref{eq5})) of height amounting the QR-surrounding matrix band offset ($V_c$ in (\ref{eq5})). Panel (c) in the same figure shows the results obtained in the case of a wider barrier($a=0.003$) with the same height. Finally, panels (a) and (b) correspond to the wider barrier but with heights amounting 10\% and 50\% of the band offset. Thus, in case (a) the barrier height (26 meV) is much smaller than the energy of the ground state of the barrier-less QR (107 meV) while in case (b) it has a height of 161 meV, larger than the energy of the nine low-lying states of the barrier-less QR (see first panel of figure \ref{Fig4}).\\

A first relevant difference with respect to the case of eccentric QRs, studied in the previous section, is that while tuning eccentricity one may separate pairs of low-lying states from the rest of the spectrum (which is little sensitive to eccentricity), a small double barrier affects all the levels to a similar extent. Differences in barriers height or width translate into more or less severe anti-crossings, but these have similar strength for all low-lying states (see figure \ref{Fig5}).\\

In order to gain insight on this issue, let us consider a limit case of a one-dimensional QR at zero magnetic field. Degenerate pairs of states are defined by wave functions $\Psi_{\pm}=e^{\pm i m \theta}$, or equivalently, by $\sin m \theta$ and $\cos m \theta$. We take the real functions and consider a perturbation defined by

\begin{equation}
\label{eq6}
V(\theta)=\left\{ \begin{array}{ll}
V_c & {\rm if} \;\;\;  -\frac{\Delta \theta}{2}< \theta <\frac{\Delta \theta}{2} \;\;\; ; \Delta \theta\; {\rm small}\\
V_c & {\rm if} \;\;\; \pi-\frac{\Delta \theta}{2}< \theta <\pi+\frac{\Delta \theta}{2} \;\;\; ;\Delta \theta\; {\rm small}\\
0 & {\rm otherwise}. \end{array} \right.
\end{equation}

By using first order perturbation theory for degenerate states, and taking into account that, for symmetry reasons, $\int_0^{2 \pi} \sin m \theta \; V(\theta) \; \cos m \theta \; d\theta =0$, the splitting of the degenerate states produced by $V(\theta)$ for very small $\Delta \theta$ is given by the difference between $\int_0^{2 \pi} \cos m \theta \; V(\theta) \;  \cos m \theta \; d\theta \approx V_c \; \Delta \theta$ and $\int_0^{2 \pi} \sin m \theta \; V(\theta) \;  \sin m \theta \;  d\theta \approx 0$, which is $m$-independent. This analysis helps us to understand why, for a quite large range of barrier heights and widths, the effect of a double barrier is to split the degenerated levels to a similar extent. \\

\begin{figure}[p]
\begin{center}
\includegraphics[width=0.7\columnwidth,angle=270]{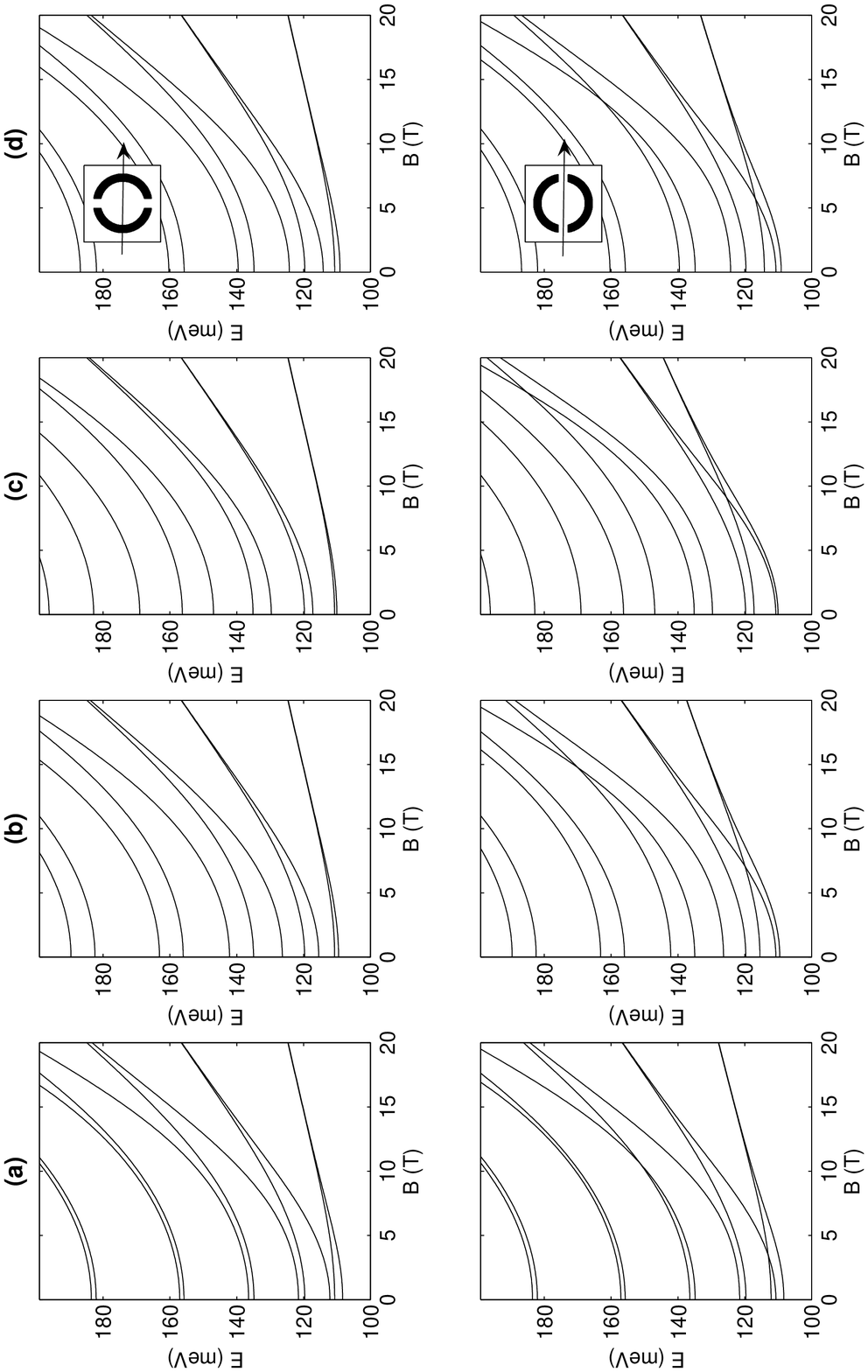}
\caption{Same as figure \ref{Fig5} but for in-plane magnetic fields. The direction of the field with 
respect to the barrier is the same for all panels in a row, 
and it is represented by the arrow in the inset.}\label{Fig6}
\end{center}
\end{figure}

The application of an axial magnetic field (see figure \ref{Fig5}) reproduces periodically the energy level positions encountered at zero magnetic field (Aharonov-Bhom effect). The case of in-plane magnetic field transverse to the double barrier (figure \ref{Fig6}, upper panels), except for the splitting of the degenerate levels at $B=0$T, which is now present for all low-lying states, resembles that of in-plane magnetic field aligned with the larger radius of an eccentric QR, for a non-severe eccentricity (figure \ref{Fig2}, upper panels). Conversely, an in-plane magnetic field aligned with the double barrier (figure \ref{Fig6}, lower panels) parallels the case of  an in-plane magnetic field transverse to the eccentric QR larger radius (figure \ref{Fig2}, lower panels). The underlying reason is that eccentricity and parallel double barriers produce opposite effects on the electronic density distribution. While eccentricity yields a larger room for the electronic density to spread along the larger radius direction of a eccentric QR, the double barrier hinders its distribution.
As in the case of geometric deformations, discussed in the previous subsection, the response of the electron energy levels to the in-plane magnetic field may be used to obtain spectroscopical information on the direction of anisotropy of the QR.

\subsection{Random imperfections in QRs}

In this section we simulate random imperfections by modifying the barrier-less reference QR with an incomplete barrier of height amounting the QR-surrounding matrix band offset, a fixed width ($a=0.003$), and length $d= 20\%, 40\%$ and $70\%$ of the QR width. This incomplete barrier is defined as:

\begin{equation}
\label{eq7}
V(\alpha,\beta)=\left\{ \begin{array}{ll}
V_c \; e^{-a \beta^2}& {\rm if} \;\;\;  r_{in}< \alpha <(r_{out}-d \; \frac{r_{out}-r_{in}}{100}), \\
0 & {\rm otherwise}, \end{array} \right.
\end{equation}
\noindent where, $\alpha=x$ and $\beta=y$ for a barrier aligned with the magnetic field, and opposite, i.e., $\alpha=y$ and $\beta=x$, if magnetic field and barrier are transverse.\\

\begin{figure}[p]
\begin{center}
\includegraphics[width=0.7\columnwidth,angle=270]{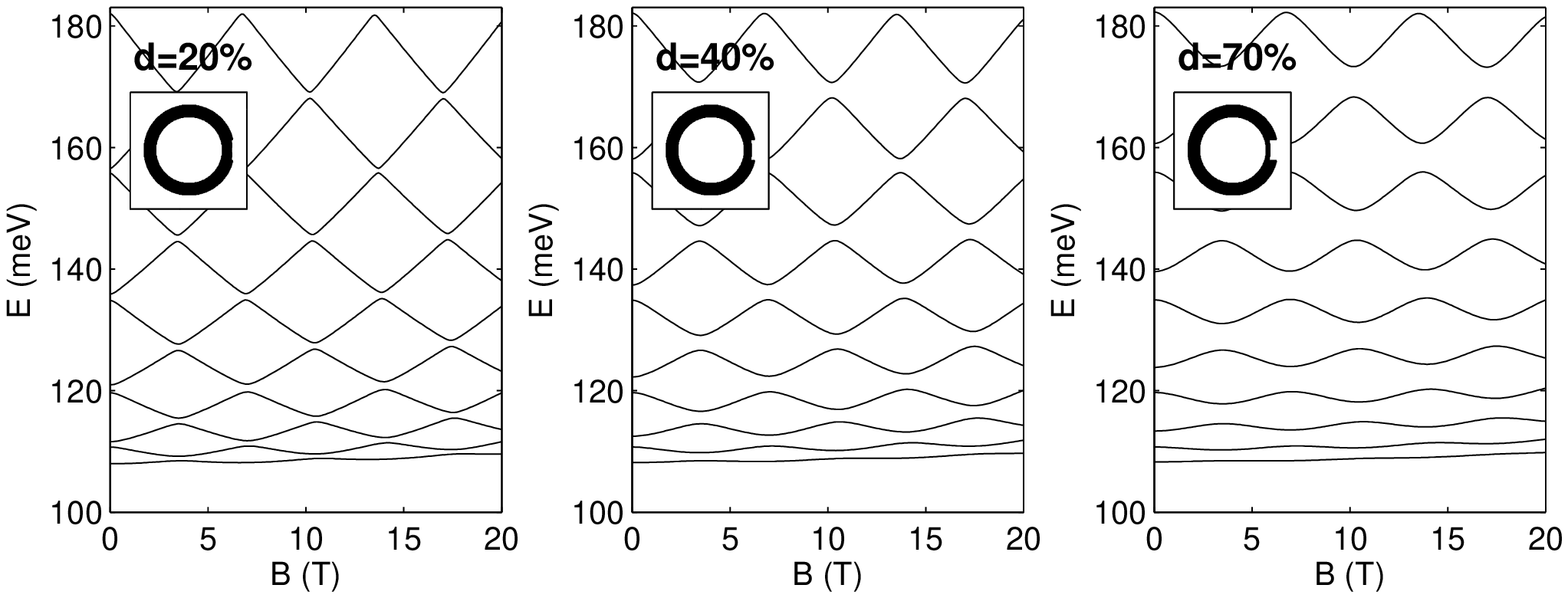}
\caption{Electron low-lying energy levels in QRs with a single-barrier structure, subject to axial magnetic fields.
The insets schematically depict the QRs shape. The barrier blocks the QR arm up to a percentage $d$ of the ring 
width.}\label{Fig7}
\end{center}
\end{figure}

The results are quite similar to those of a parallel double barrier QR, except for the additional anti-crossings in the presence of an axial magnetic field, related to the reduction of symmetry from $C_2$ corresponding to a parallel double barrier QR up to $C_1$ of a single barrier QR. It can be made apparent by comparing Figs.~\ref{Fig5} and \ref{Fig7}. In the case of an incomplete barrier, we additionally observe that the larger the length $d$ is the wider the anti-crossings result (see figure \ref{Fig7}).\\

\begin{figure}[p]
\begin{center}
\includegraphics[width=0.7\columnwidth,angle=270]{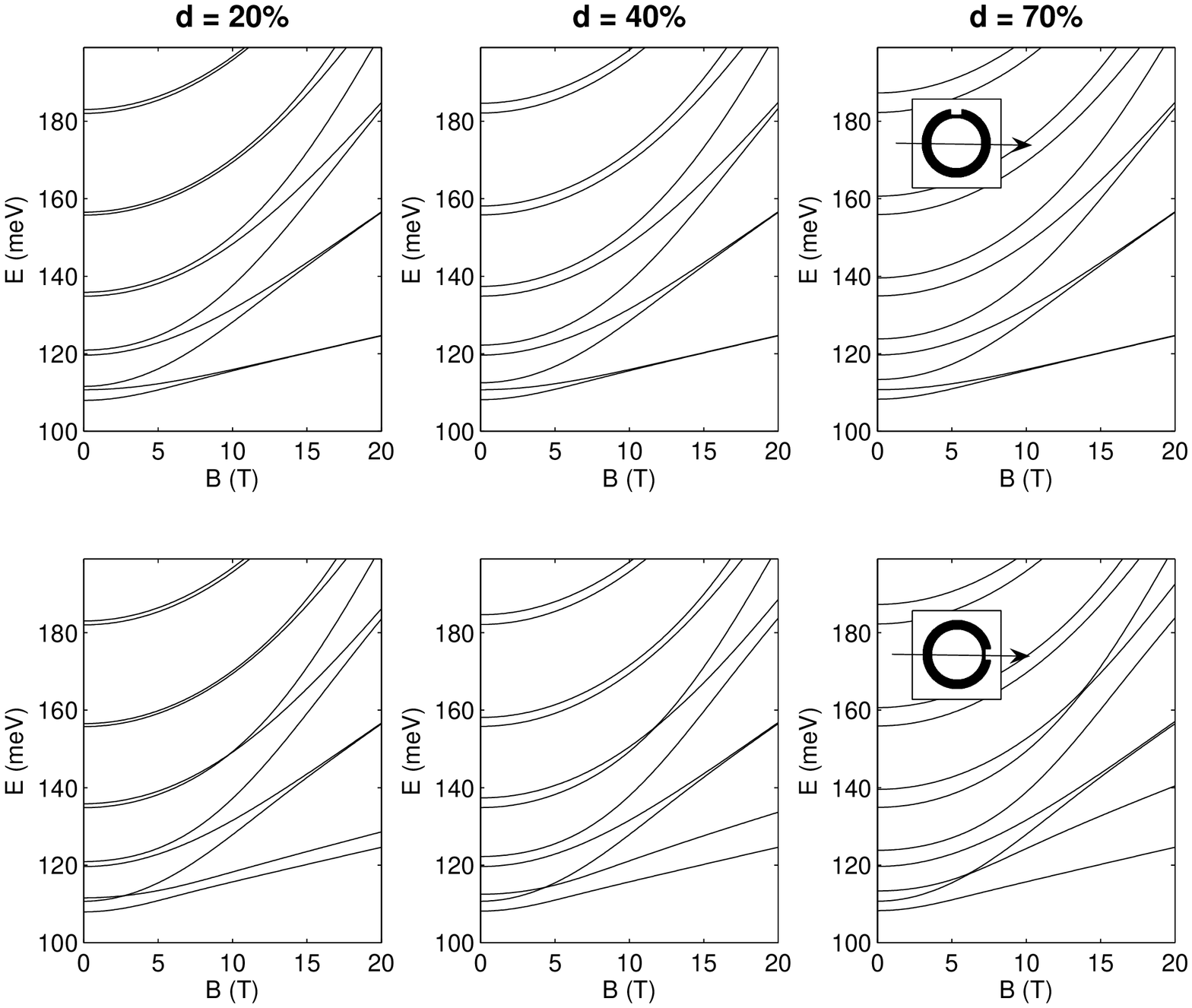}
\caption{Same as figure \ref{Fig7} but for in-plane magnetic fields. The direction of the field with 
respect to the barrier is the same for all panels in a row, 
and it is represented by the arrow in the inset.}\label{Fig8}
\end{center}
\end{figure}

The in-plane magnetic field, both, aligned with and transverse to the barrier, produces energy magnetospectra qualitatively similar to those of a parallel double barrier QR subject to the same fields (it is apparent by comparing Figs.~\ref{Fig6} and \ref{Fig8}). The underlying reason is that, at zero magnetic field, no degenerate states occur in both systems, which can be in turn qualitatively modeled by the perturbation produced on the pair of degenerate wave functions ($\sin m \theta$, $\cos m \theta$) of a barrier-less QR by either, $V(\theta)$ given by (\ref{eq6}), that simulates the parallel double barrier QR case, or the perturbation produced by, 

\begin{equation}
\label{eq7b}
V(\theta)=\left\{ \begin{array}{ll}
V_c & {\rm if} \;\;\;  -\frac{\Delta \theta}{2}< \theta <\frac{\Delta \theta}{2} \;\;\; ; \Delta \theta\; {\rm small}\\
0 & {\rm otherwise}, \end{array} \right. 
\end{equation}

\noindent which mimics the case of a single barrier QR. When the magnetic field is switched on axially, differences arise coming from a distinc system symmetry, either leaving a crossing ($C_2$) or only yielding anti-crossings ($C_1$). However, in the presence of an in-plane magnetic field, both systems have the same symmetry and yield qualitatively similar energy magnetospectra.

\section{Concluding Remarks}
The electron energy level magneto-spectra of QR with defects are investigated. Several models are employed to mimic geometry anisotropic deformation and mass diffusion occuring in the QR fabrication process. The magnetic field is applied axial and in-plane including both directions, aligned and transverse to the anisotropic defect. The electron energy levels of anisotropic rings show a clearly different response to longitudinal and transverse in-plane magnetic fields, which may allow to estimate spectroscopically the direction and the extension of the deformation. Physical insight on the obtained numerical results is given by symmetry considerations and simplified analytical models. We conclude that, except for unusual extremely large eccentricities, QR geometry deformations only appreciably influence a few low-lying states, while mass diffusion, modeled with barriers disturbing the QR persistent current, has on the one hand a larger effect on the energy levels and, on the other hand it affects all studied states to a similar extent.

\ack
Continuous support from MEC-DGI projects CTQ2004-02315/BQU, and UJI-Bancaixa project P1-1B2006-03 are gratefully 
acknowledged.  One of us (J.I.C.) has been supported by the EU under the Marie Curie IEF project
MEIF-CT-2006-023797.



\begin{thebibliography}{50}
\bibitem{note1} The Coulomb gauge does not determine $\mathbf{A_{axial}}$ completely, we could select  $\mathbf{A'_{axial}}=(0,x,0) B$ instead.
\bibitem{note2} The Coulomb gauge does not determine $\mathbf{A_{in}}$ completely. The employed gauge, $\mathbf{A_{in}}=(0,0,y) B$, allows to carry out an effective separation of $z$ and $(x,y)$ variables in the Hamiltonian (\ref{eq1}), which is crucial to study a flat in-plane system.\cite{planellesPE} Note, though, that using a (bonafide) gauge $\mathbf{A} = (0,-z,0) B$ we would not be able to make this effective separation of variables.
\bibitem{degen} Each single low-lying line appearing in figure \ref{Fig1} for $e=0.60,\, 0.80$ and $0.95$ is, actually, a superposition of two lines, corresponding to two almost degenerate states.
\bibitem{accidental} The degeneracy of the fourth and fifth pairs of energy levels in the $e=0.95$ magneto-spectrum of figure \ref{Fig1} is accidental.
\bibitem{e-e} If the QR is populated with two or more electrons, the many-body energy spectrum will display similar periodic crossings and anticrossings to those we describe here but with a shorter oscillation period. On the one hand, Coloumb interactions are responsible of this shortening of the oscillation periods (fractional Aharanov-Bohm effect \cite{NiemelaEPL,PlanellesBGW}). On the other hand, anti-crossings are related to the reduction of symmetry from $C_{\infty}$ up to $C_n$, produced by the barriers (note that Coloumb interaction does not affect the point symmetry group of the Hamiltonian, because it depends on the inter-electronic distance only and symmetry operations do not change  distances).
\end{thebibliography}
\end{document}